# Common data file definitions for neutron inelastic scattering instruments using NeXus

Jörn Beckmann, Peter Link, Tobias Unruh, Nils Pyka, Dirk Etzdorf

Technische Universität München
ZWE FRM-II
Garching, Germany
jbeckman@frm2.tum.de

The NeXus format allows the storage of measured data next to a description of sample and instrument setup in a standardized manner. The NeXus standard itself gives only general rules how an instrument setup is described. A so called instrument definition contains a minimal set of information and the location of that information within a file.
In this paper instrument definitions for neutron triple-axis and time-of-flight reactor machines are suggested. Furthermore a structure for the storage of instrument logbook information is presented.

PRESENTED AT

NOBUGS 2002
Gaithersburg, MD, U.S.A, November 3-5, 2002

# 1 Introduction to NeXus data Format

NeXus is a structured data format developed by leading neutron scattering centers and supported by major facilities. The NeXus standard is implemented in the NeXus API which operates upon the HDF format. Software which supports HDF is also able to operate on files created by the NeXus API. Versions of the NeXus API are currently available for C, Fortran 77, Fortran 90, Java and IDL. Details about the NeXus standard and API can be found on the NeXus web page [1].

A NeXus file made up of NeXus objects which could either be data items or groups. A group is like a folder containing subgroups and data items giving a NeXus file the overall appearance of a tree structure or a file system. A NeXus group has a name and belongs to a class. There can be more than one group of the same class within a NeXus file as long as the names of the groups are different. Each NeXus object could hold extra information like units in the form of attributes. The generic structure of a NeXus file is given in figure 1.

The root of the NeXus structure tree is NXentry which contains all other groups like NXsample (describing the sample investigated), NXinstrument (giving a description of the actual instrument setup) and NXdata (holding the actual data). The NXdata group contains an array for each data axis, an array with the values of the signal and an optional array of error values. The different arrays are distinguished by attributes. Either each axis contains an "axis" attribute giving the rank of the different axes and the data object contains a "signal" attribute or the data object just contains an "axes" attribute which is an array of the names of the different axes.

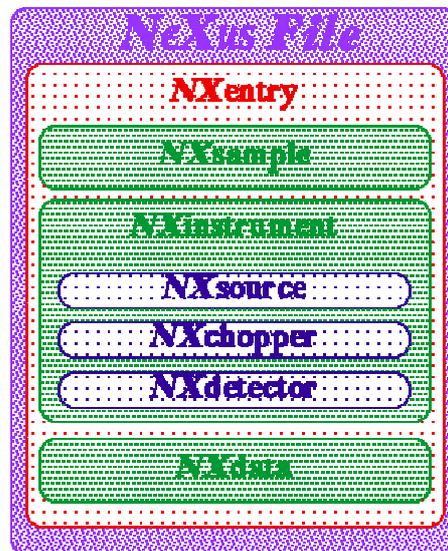

Figure 1: Structure of a NeXus file.

# 2 Nomenclature and ambiguity of existing definitions

The NeXus format outlined above just defines how information is stored within the NeXus file. In the NXinstrument group, for example, all distances are given relative to the sample, negative distances indicate that the component is between source and sample position. But the standard does not say if the distance is measured from the middle of the sample or from the point next to the sample and whether the distance is measured along the beam. A second open point is the orientation in which angels are given. While this is no problem as long as files from different instruments are kept separately but it becomes a major issue for a triple-axis instrument definition.

Instrument definitions are given in XML using the so called Meta-DTD definition which could be found in [2]. The format is designed in a way that an instrument definition could be created from a NeXus file by converting it to XML and omitting the data, which is done by the NXtoDTD tool. This makes the instrument definition files easy readable but it prevents automatic verification of NeXus files against an instrument definition. Therefore in addition to the Meta-DTD a real DTD in standard format should be provided to allow the usage of XML verification tools.

In the Meta-DTD after an opening tag a symbol indicates if the entry may occur one or more times and if it is mandatory. The symbols are * (entry may occur 0 or more times), + (entry must occur at least once) and ? (entry may occur 0 or one time). If no symbol is given the entry is mandatory. Lines given in green color indicate objects added to groups defined in the NeXus standard. The structure of NeXus groups not yet defined in the standard or significantly changed are presented in chapter 5. Groups which are already foreseen in the standard are named starting with "NX" while groups special to FRM-II needs start with "FRM" in the name.

# 3 Instrument definition for a Triple-Axis-Spectrometern (TAS)

The following instrument definition is made up from the definitions of the instruments PANDA [3] and PUMA [4] at the FRM-II reactor. PANDA is a cold polarized TAS, while PUMA is a thermal TAS with multianalyzer option. The file is constructed in a way, that scans of different parameters with the same sample could be stored within one NXentry in different NXdata groups. If the geometry setup of the experiment has to be changed between scans an additional NXentry in the same NXfile must be created. The NXmonitor group was moved from NXentry into the NXinstrument subgroup as it seems to be more consistent to keep all instrument parts together in one group.

For each datapoint in a scan entries have to be created within several structures. The idea is to store all parameters which might change during a scan in arrays. For each datapoint there is an entry in every array even if the value didn't change. The datapoint array in NXdata keeps track of the datapoints of a single scan. While the datapoint array is local for each scan with an own NXdata group, the global arrays like temperature in NXsample are shared by all scans within one NXentry. The offset attribute of the datapoint object indicates the beginning of the scan in the global arrays. For example: The NXentry contains 80 datapoints of two scans. Datapoints 1 to 40 belong to a ki scan, while datapoint 41 to 80 belong to a temperature scan. The NXentry contains two NXdata groups with datapoint arrays of 40 entries each. The temperature object in the NXsample group and the ki array in the FRM_scattering_triangle group contain 80 points each. The offset attribute of the datapoint array in the first NXdata group (belonging to the ki scan) is set to "0", while it is set to "41" in the second NXdata group (belonging to the temperature scan). The scanvariable array in the first NXdata group therefore holds ki values while the scanvariable array of the second NXdata group holds temperature values. It is the task of the data processing software to keep track on that. To ease the creation of generic plotting tools datapoint was chosen as the primary axis of the NXdata group. The full instrument definition is given below:

<?xml version="1.0" ?>

```
<!--
URL:     http://www.neutron.anl.gov/nexus/xml/NXntas.xml
Editor:  Joern Beckmann <jbeckman@frm2.tum.de>
$Id$
Instrument definition for a neutron triple-axis spectrometer.
-->
<NXfile file_time="{Time when this file was originally created}" file_name="{Name of this file when
originally created}" NeXus_version="{Version of NeXus API}">
          <NXentry name="{Name of measurement}">+
                  <title>{Extended title for entry}</title>
                  <analysis version="$Revision$"
                          URL="http://www.neutron.anl.gov/nexus/xml/NXntas.xml">
                          NTAS
                  </analysis>
                  <notes>? {Notes describing the entry}</notes>
                  <NXuser  name="{Username or ID}">
                          <user_id type="NX_INT32">? {ID of user at installations DB}</user_id>
                          <name> {Name of scientist responsible} </name>
                          <affiliation>? {Institution of user} </affiliation>
                          <address>? {users address} </address>
                          <telephone_number>? {Phone number} </telephone_number>
                          <fax_number>? (Fax number) </fax_number>
                          <email>? {Email address} </email>
                  </NXuser>
                  <FRMlocalcontact name="{Person responsible at installation}">?
                  <NXsample name="{ID of sample}">?
                          <name> {Descriptive name} </name>
                          <chemical_formula>? {Composition of sample} </chemical_formula>
                          <environment>? {Description of sample environment}  </environment>
                          <temperature type="NX_FLOAT32[i]" name="{Description where
                                  temperature is measured}">* {Temperature values}
                          </temperature>
                          <electric_field type="NX_FLOAT32[3,i]">? {Applied field vector}
                          </electric_field>
                          <magnetic_field type="NX_FLOAT32[3,i]">? {Applied field vector}
                          </magnetic_field>
                          <stress_field type="NX_FLOAT32[3,i]">? {Applied stress} </stress_field>
                          <pressure type="NX_FLOAT32[i]">? {Applied pressure} </pressure>
                          <changer_position type="NX_INT32">? {Sample changer position}
                          </changer_position>
                          <unit_cell type="NX_FLOAT32[6]">? {Unit cell length and angles}
                          </unit_cell>
                          <unit_cell_volume unit="Angstroms^3" type="NX_FLOAT32">?
                                  {Volume of unit cell} </unit_cell_volume>
                          <orientation_matrix type="NX_FLOAT32[9]">? {Orientation of crystal
                                  sample} </orientation_matrix>
                          <symmetry_cell_setting>? "cubic" | "tetragonal" | "hexagonal" |
                                  "rhomboedric" | "orthorhombic" | "monoclinic" | "triclinic"
                          </symmetry_cell_settings>
                          <shape>? "plate" | "sphere" | "cylinder" | "hollow cylinder" </shape>
                          <dimension units="cm" type="NX_FLOAT32[3]">? {Dimension of plate-
                                  like sample} </dimension>
                          <radius units="cm" type="NX_FLOAT32">? {Radius of cylindrical or
                                  spherical sample} </radius>
                          <inner_radius units="cm" type="NX_FLOAT32">? {Inner radius of
                                  hollow sample}</inner_radius>
                          <height units="cm" type="NX_FLOAT32">? {Height of spherical
                                  sample}</height>
                          <mass units="g" type="NX_FLOAT32">? {Mass of sample}</mass>
                          <density units="g cm-3" type="NX_FLOAT32">? {Density of sample}
                          </density>
```

```
<molecular_weight units="g mol-1" type="NX_FLOAT32">? {Molecular
        weight of sample}</molecular_weight>
<coherent_cross_section units="barns" type="NX_FLOAT32" >?
        {Coherent cross section} </coherent_cross_section>
<incoherent_cross_section units="barns" type="NX_FLOAT32">?
        {Incoherent cross section} </incoherent_cross_section>
<absorption_cross_section units="barns" type="NX_FLOAT32">?
        {Absorption cross section} </absorption_cross_section>
<NXlog name="{Sample property logged}">*
        <value type="NX_FLOAT32[:]" {Value of variable being
                logged} </value>
        <time_log type="ISO 8601[:]"> {Time at which variable is
                logged} </time_log>
</NXlog>
</NXsample>
<NXinstrument name="{Name of instrument}">
        <NXsource name="{Name of source}">+
                <distance units="m" type="NX_FLOAT32"> {Distance from
                        sample along the beam} </distance>
                <type> "Pulsed Reactor" | "Reactor" </type>
                <power units="MW" type="NX_FLOAT32"> {Source power}
                        </power>
                <frequency units="Hz" type="NX_FLOAT32">? {frequency of
                        pulsed source} </frequency>
                <period units="microseconds" type="NX_FLOAT32">? {Period
                        of pulsed source} </period>
                <moderator>? "H2O" | "D2O" | "H2" | "D2" | "C" </moderator>
                <moderator_temperature units="Kelvin" type="NX_FLOAT32">
                        ? {Temperature of moderator} </moderator_temperature>
                <fuel_element_id>? {Fuel element or reactor cycle}
                        </fuel_element_id>
        </NXsource>
        <NXaperture>*
        <NXchopper name="{Name of chopper}">*
                <distance units="m" type="NX_FLOAT32"> {Distance from
                        sample along the beam} </distance>
                <type>? "Fermi" | "Disk" | "Counter rotating disk" | "Velocity
                        selector" </type>
                <frequency units="Hz" type="NX_FLOAT32">? {Rotation
                        speed} </frequency>
                <period units="microseconds" type="NX_FLOAT32">? {Period
                        of chopper rotation} </period>
                <radius units="cm" type="NX_FLOAT32">? {Radius of chopper
                        body} </radius>
                <curvature units="cm" type="NX_FLOAT32">? {Curvature of
                        Fermi chopper} </curvature>
                <slit_width units="cm" type="NX_FLOAT32">? {Width of
                        Fermi chopper slits} </slit_width>
                <slit_number type="NX_INT32">? {Number of Fermi chopper
                        slits} </slit_number>
                <blade_width units="cm" type="NX_FLOAT32">? {Width of
                        Fermi chopper blades} </blade_width>
                <energy calibration_status="'Nominal' | 'Measured'" units="meV"
                        type="NX_FLOAT32[i]">? {Optimum energy
                        transmitted by he chopper} </energy>
                <transmission type="NX_FLOAT32[i]">? {Transmission of the
                        device} </transmission>
                <resolution type="NX_FLOAT[:]">? {Energy resolution of the
                        device} </resolution>
                <mode>? "connected" | "disconnected" </mode>
                <direction>? "clockwise" | "counter clockwise"</direction>
```

```
                <NXlog name="Trigger Log">?
                        <value type="NX_FLOAT32[:]"> {Trigger pulses}
                        </value>
                        <time_log type="ISO 8601[:]"> {Timestamp}
                        </time_log>
                </NXlog>
                <NXlog name="Phase Log">?
                        <value type="NX_FLOAT32[:]"> {Chopper phase}
                </value>
                        <time_log type="ISO 8601[:]"> {Timestamp}
                        </time_log>
                </NXlog>
                <NXlog name="Frequency Log">?
                        <value type="NX_FLOAT32[:]"> {Chopper frequency}
                        </value>
                        <time_log type="ISO 8601[:]"> {Timestamp}
                        </time_log>
                </NXlog>
        </NXchopper>
        <NXmirror>*
        <NXcollimator name="{Description of collimator}">*
                <distance units="m" type="NX_FLOAT32"> {Distance to sample
                        along the beam} </distance>
                <type> "Soller" </type>
                <length units="cm" type ="NX_FLOAT32"> {Length of
                        collimator} </length>
                <soller_angle units="minutes" type="NX_FLOAT32">?
                        {Angular divergence of Soller collimator}
                </soller_angle>
                <horizontal_aperture units="cm" type="NX_FLOAT32">?
                        {Horizontal aperture of collimator, if rectangular}
                </horizontal_aperture>
                <vertical_aperture units="cm" type="NX_FLOAT32">? {Vertical
                        aperture of collimator, if rectangular}
                </vertical_aperture>
                <radius units="cm" type="NX_FLOAT32">? {Aperture radius of
                        collimator, if circular} </radius>
                <transmission type="NX_FLOAT32">? [Transmission of
                        collimator} </transmission>
        </NXcollimator>
        <NXcrystal name="{Name of crystal beamline component}">+
                <distance units="cm" type="NX_FLOAT32"> {Distance from
                        sample along the beam} </distance>
                <wavelength units="Angstroms" type="NX_FLOAT32[i]">?
                        {Optimal reflected wavelength} </wavelength>
                <energy units="meV" type=NX_FLOAT32[i]>? {Nominal
                        energy} </energy>
                <wavevector units="1/Angstroms" type="NX_FLOAT32[i]">?
                        {Nominal wavevector} </wavevector>
                <lattice_parameter units="Angstroms" type="NX_FLOAT32">?
                        {Lattice parameter of nominal reflection}
                </lattice_parameter>
                <reflection type="NX_FLOAT32[3]"> {hkl values of nominal
                        reflection} </reflection>
                <horizontal_curvature units="degrees"
                        type="NX_FLOAT32[i]">? {Horizontal curvature of
                        focusing crystal} </horizontal_aperture>
                <vertical_curvature units="degrees" type="NX_FLOAT32[i]">?
                        {Vertical curvature of focusing crystal}
                </vertical_curvature>
                <horizontal_aperture units="cm" type="NX_FLOAT32[i]">?
```

```xml
                        {Horizontal size of aperture, if rectangular}
                    </horizontal_aperture>
                    <vertical_aperture units="cm" type="NX_FLOAT32[i]">?
                        {Vertical size of aperture, if rectangular}
                    </vertical_aperture>
                    <radius units="cm" type="NX_FLOAT32[i]">? {Radius of
                        aperture, if circular} </radius>
            </NXcrystal>
            <NXflipper>?
            <NXdetector name="{Name of detector}">+
                    <distance units="m" type="NX_FLOAT32"> {Distance from
                        sample along the beam} </distance>
                    <id type="NX_INT32">? {Identifier of detector element}</id>
                    <type ?"He3 gas cylinder" | "He3 PSD" | "He3 multidetector" |
                        "BF3 gas cylinder" | "scintillator" | "fission chamber"
                    </type>
                    <gas_pressure units="bars" type="NX_FLOAT32">? {Detector
                        gas pressure} </gas_pressure>
                    <efficiency type="NX_FLOAT32[:]">? {Efficiency of detector}
                    </efficiency>
                    <height units="cm" type="NX_FLOAT32">? {Height of detector
                        element} </height>
                    <width units="cm" type="NX_FLOAT32">? {Width of detector
                        element} </width>
                    <depth units="cm" type="NX_FLOAT32">? {Depth of detector
                        element} </depth>
                    <radius  units="cm" type="NX_FLOAT32">? {Radius of detector
                        element} </radius>
            </NXdetector>
            <NXmonitor name="{Description of monitor}">*
                    <distance units="m" type="NX_FLOAT32"> {Distance from
                        sample along the beam} </distance>
                    <type>? "fission chamber" | "Scintillator" </type>
                    <height units="cm" type="NX_FLOAT32">? {Height of
                        monitor} </height>
                    <width units="cm" type="NX_FLOAT32">? [Width of monitor]
                    </width>
                    <efficiency type="NX_FLOAT32[:]">? {Monitor efficiency}
                    </efficiency>
                    <data units="counts" type="NX_INT32[i]"> {Monitor counts}
                    </data>
                    <NXlog name="Monitor Log">
                            <value type="NX_INT32[:]"> {Monitor rate} </value>
                            <time_log type="ISO 8601[:]">{Timestamp}
                            </time_log>
                    </NXlog>
            </NXmonitor>
    </NXinstrument>
    <FRM_Scattering_triangle>?
    <FRM_Scattering_angles>?
    <NXdata scantype="{Description of scan performed}">+
            <datapoint axis="1" primary="1" offset="{Index of the first local
                    datapoint element in the global arrays}" type="NX_INT32[j]">
                {Unique datapoint ID} </datapoint>
            <scanvariable axis="2" type="NX_FLOAT32[j]">? {Scan parameter like
                    angle or temperature} </scanvariable>
            <counts source="name of detector" signal="1" type="NX_INT32[j]">+
                {Detector counts} </counts>
            <errors source="name of detector" type="NX_INT32[j]">* {Error bar of
                    detector counts} </errors>
    </NXdata>
```

```
            <FRM_CommandLog>?
        </NXentry>
</NXfile>
```

# 4    Instrument definition for a reactor time-of-flight spectrometer

While the instrument definition for TAS started from scratch the following definition
for a reactor-type time-of-flight machine is based on the NXtofndgs.xml definition [5]
and reflects features of the TOFTOF [6] instrument from FRM-II. Below one finds an
overview of the file structure. Subgroups which are not explicitly printed, have the
same structure such as in the NXntas.xml definition. The main differences to
NXtofndgs.xml are the addition of a radial collimator and a different description of
the detector arrays. The efficiency of the detectors is not given explicitly. It has to be
calculated from a formula given in the NXdetector section. The coefficients of that
formula have to be determined experimentally and are given in the same group.

```
<?xml version="1.0" ?>
<!--
URL:    http://www.neutron.anl.gov/nexus/xml/NXrtof.xml
Editor:  Joern Beckmann <jbeckman@frm2.tum.de>
$Id$
Instrument definition for a reactor-type time-of-flight spectrometer
-->
<NXfile file_time="{Time when this file was originally created}" file_name="{Name of this file when
originally created}" NeXus_version="{Version of NeXus API}">
        <NXentry name="{Name of measurement}">+
                <title>{Extended title for entry}</title>
                <analysis version="$Revision$"
                        URL="http://www.neutron.anl.gov/nexus/xml/NXrtof.xml">
                        RTOF
                </analysis>
                <notes>?
                <NXuser>
                <FRMlocalcontact>?
                <NXsample>
                <NXinstrument>
                        <NXsource>
                        <NXchopper>
                        <NXcollimator name="{Description of collimator}">*
                                        :
                                        :
                                <rotation_speed type="NX_FLOAT32">? {Speed of radial
                                        collimator} </rotation_speed>
                                <amplitude type="NX_FLOAT32">? {Amplitude of radial
                                        collimator} </amplitude>
                                <inner_radius type="NX_FLOAT32">? {Inner radius of radial
                                        collimator} </inner_radius>
                                <outer_radius type="NX_FLOAT32">? {Outer radius of radial
                                        collimator} </outer_radius>
                                <mode>? "continuous" | "discontinuous" </mode>
                        </NXcollimator>
                        <NXdetector>
                                        :
                                <solid_angle units="steradians" type=NX_FLOAT32>? {Angle
                                        between detector axis and incident beam} </solid_angle>
                                <row type="NX_INT32">? {Detector position} </row>
                                <column type="NX_INT32">? {Detector position} </column>
                                <efficiency_formula>? {Formula to calculate the efficiency of the
                                        detector at a given wavelength} </efficiency_formula>
```

```
                    <coefficient_1 type="NX_FLOAT32">? {First coefficient for
                              efficiency formula}</coefficient_1>
                    <coefficient_2 type="NX_FLOAT32">? {Second coefficient for
                              efficiency formula} </coefficient_2>
            </NXdetector>
            <NXmirror>
            <FRM_Beamstop>
            <NXmonitor>
        </NXinstrument>
        <NXdata>+
            <time_of_flight axis="1" histogram_offset="{Offset to first histogram bin
                       center}" type="NX_FLOAT32[n]> {Histogram bins}
            </time_of_flight>
            <id axis="2" type="NX_INT32[n]> {Name of detector} </id>
            <counts signal="1" type="NX_FLOAT32[n]> {Counts per bin} </counts>
        </NXdata>
        <FRM_CommandLog>?
    </NXentry>
</NXfile>
```

# 5    Proposed Extensions to the NeXus standard

As there is no NXaperture defined in the standard, yet we used the definition given below:

```
    <NXaperture name="{name of aperture}">
            <distance units="m" type="NX_FLOAT32"> {Distance from sample along the
                       beam}</distance>
            <shape>? "Rectangular" | "Circular" | "Elliptical" </shape>
            <horizontal_aperture units="cm" type="NX_FLOAT32[:]>? {Horizontal size}
            </horizontal_aperture>
            <vertical_aperture units="cm" type="NX_FLOAT32[:]>? {Vertical size}
            </vertical_aperture>
            <radius units="cm" type="NX_FLOAT32[:]>? {Radius of circular aperture}
            </radius>
    </NXaperture>
```

Not only information about the user who performed the measurement are important but also information about the instrument responsible. This information is stored in the FRMlocalcontact subgroup which is similar to the NXuser group, just affiliation and the database id are missing.

```
    <FRMlocalcontact name="{Name of local responsible}">
            <name> {Name of responsible} </name>
            <address>? {Address of responsible} </address>
            <telephone_number>? {Telephone number of responsible} </telephone_number>
            <fax_number>? {Fax number of responsible} </fax_number>
            <email>? {Email address of responsible} </email>
    </FRMlocalcontact>
```

For TAS experiments the values of hkl, the incident and the analyzed energy are of interest. The FRM_Scattering_triangle subgroup stores these values at one place and makes it easier for TAS-software to find the necessary information.

```
    <FRM_Scattering_triangle>
            <h type="NX_FLOAT32[:]>? {h value of wavevector} </h>
            <k type="NX_FLOAT32[:]>? {k value of wavevector} </k>
            <l type="NX_FLOAT32[:]>? {l value of wavevector} </l>
```

```
        <ki units="1/Angstroms" type="NX_FLOAT32[:]">? {ki value of incident
                wavevector} </ki>
        <kf units="1/Angstroms" type="NX_FLOAT32[:]">? {kf value of analyzed
                wavevector} </kf>
</FRM_Scattering_triangle>
```

To be sure whether the desired values of FRM_Scattering_triangle were really set during the experiment, the values of the physical instrument settings are needed. FRM_Scattering_angles keeps them together. But these values are instrument specific and not suitable to be processed by generic TAS-software.

```
<FRM_Scattering_angles>
        <theta_M units="degrees" type="NX_FLOAT32[:]">? {theta angle of
                monochromator} </theta_M>
        <twotheta_M units="degrees" type="NX_FLOAT32[:]">? {twotheta value of
                monochromator} </twotheta_M>
        <theta_A units="degrees" type="NX_FLOAT32[:]">? {theta value of analyzer}
                </theta_A>
        <twotheta_A units="degrees" type="NX_FLOAT32[:]">? {twotheta value of
                analyzer} </twotheta_A>
        <omega units="degrees" type="NX_FLOAT32[:]">? {omega value of sample
                orientation} </omega>
        <phi units="degrees" type="NX_FLOAT32[:]">? {phi value of sample orientation}
                </phi>
<FRM_Scattering_angles>
```

When users discover strange results during data processing they like to know whether during the experiment all hardware worked as desired or if there was an error in the scanning script. In such a situation it is helpful to have the full command history of the experiment including all error messages.

```
<FRM_Command_Log>
        <command type="NX_CHAR[:]"> {Commands entered} </command>
        <reply type="NX_CHAR[:]"> {Reply of Command} </reply>
        <timestamp type="ISO 8601 [:]"> {Time when command was submitted}
                </timestamp>
</FRM_Command_Log>
```

Definition for a mirror guide, flipper or a beamstop are not yet included in the standard. Our definitions are given below:

```
<NXmirror name="{ID of guide}" type="'polarizing' | 'non-polarizing'">
        <distance units="m" type="NX_FLOAT32" {Distance from middle of guide to
                sample along the beam} </distance>
        <length units="m" type="NX_FLOAT32"> {Length of guide} </length>
        <height units="cm" type="NX_FLOAT32" position="'front' | 'middle' | 'back'">+
                {Height of mirror guide at indicated position} </height>
        <width units="cm" type="NX_FLOAT32" position="'front' | 'middle' | 'back'">+
                {Width of mirror guide at indicated position} </width>
        <curvature units="cm" type="NX_FLOAT32">?{Radius of curvature of mirror
                guide} </curvature>
        <reflectivity type="NX_FLOAT32" orientation="'horizontal' | 'vertical'">*
                {Reflectivity of indicated guide surface} </reflectivity>
        <coating orientation="'horizontal' | 'vertical'">* {Description of surface coating}
                </coating>
</NXmirror>

<NXflipper name="{Description of flipper device}">
```

```
                <distance units="cm" type="NX_FLOAT32"> {Distance from middle of guide to
                        sample along the beam} </distance>
                <orientation>? "horizontal" | "vertical" </orientation>
                <current units="mA" type="NX_FLOAT32[:]">? {Current and polarity in flipper
                        coil}</current>
                <efficiency type="NX_FLOAT32[:]">? {Efficiency of flipper device} </efficiency>
                <NXlog name="Current Log">?
                        <value type="NX_FLOAT32[:]"> {Current in flipper coil}</value>
                        <time_log type="ISO 8601[:]"> {Timestamp}</time_log>
                </NXlog>
        </NXflipper>

        <FRM_Beamstop name="{Description of beamstop}">
                <distance units="m" type="NX_FLOAT32"> {Distance from sample along the
                        beam} </distance>
                <width units="cm" type="NX_FLOAT32">? {Width of beamstop, if rectangular}
                </width>
                <height units="cm" type="NX_FLOAT32">? {Height of beamstop, if rectangular}
                </height>
                <radius units="cm" type="NX_FLOAT32">? {Radius of beamstop, if
                        circular}</radius>
                <horizontal_position type="NX_FLOAT32"> {Horizontal position of the
                        beamstop relative to center of beam} </horizontal_position>
                <vertical_position type="NX_FLOAT32"> {Vertical position of the beamstop
                        relative to the center of beam} </vertical_position>
        </FRM_Beamstop>
```